\documentclass[conference,a4paper]{IEEEtran}
\usepackage{amsmath,graphicx,hyperref}

\usepackage[utf8]{inputenc} % allow utf-8 input
\usepackage[T1]{fontenc}    % use 8-bit T1 fonts
\usepackage{url}            % simple URL typesetting
\usepackage{booktabs}       % professional-quality tables
\usepackage{amsfonts}       % blackboard math symbols
\usepackage{nicefrac}       % compact symbols for 1/2, etc.
\usepackage{microtype}      % microtypography
\usepackage{xcolor}         % colors

\usepackage{todonotes}
\usepackage{multirow}
\usepackage{amsmath}
\usepackage{float}          % For precise table positioning with [H]

\usepackage{algorithm}

\usepackage[caption=false,font=footnotesize]{subfig}
\usepackage{booktabs}

\usepackage{balance}
\usepackage{cite}
\usepackage{amsmath,amssymb,amsfonts}
\usepackage{textcomp}
\usepackage{xcolor}
\usepackage{booktabs}
\usepackage{makecell}
\usepackage{algpseudocode}
\usepackage{multirow}
\usepackage{algorithm}
\usepackage{soul}
\usepackage{graphicx}

\usepackage{subcaption}

\title{\huge C-MambaPose: Phase-Preserving Complex Mamba for Cross-Environment Wi-Fi Human Pose Estimation}

\author{
    \IEEEauthorblockN{
        Phuc H. Nguyen$^1$ 
        and Van-Dinh Nguyen$^{2}$}
    \IEEEauthorblockA{\it $^1$College of Engineering \& Computer Science, VinUniversity, Vietnam \\
     $^2$School of Computer Science and Statistics, Trinity College Dublin,  Ireland\\
    E-mails: 23phuc.nh@vinuni.edu.vn, dinh.nguyen@tcd.ie
    \vspace{-5pt}}
    \thanks{This work was supported by the VinUniversity project CECS.UNI.2026.DA012}

}

\begin{document}
\maketitle

\begin{abstract}
Cross-environment generalization remains a fundamental challenge for Wi-Fi-based human pose estimation (HPE), as environmental multipath can drastically alter Channel State Information (CSI) for the same human motion. Existing approaches often discard CSI phase or collapse complex signals into real-valued features early, losing motion-sensitive phase dynamics. We present C-MambaPose, a phase-preserving complex-valued state-space framework for robust cross-environment Wi-Fi 3D HPE. Our key insight is to preserve complex phase relationships throughout spatiotemporal modeling, rather than using phase only as an input feature. C-MambaPose sanitizes synchronization-induced phase errors and employs a Spatiotemporal Complex Mamba encoder with dynamic multi-scale receptive fields to capture long-range phase-amplitude dynamics. Cross-attention then maps the learned representations to human joints, followed by a skeleton-aware GraFormer decoder for anatomically coherent pose estimation. Experiments on MM-Fi show that C-MambaPose achieves strong performance across all settings, with the largest gains in unseen environments, reaching $298.5$ mm and $317.6$ mm MPJPE on the two cross-environment protocols using only $3.78$ M parameters and no pretraining. Ablations further show that prematurely discarding complex phase relationships substantially degrades performance, demonstrating their importance for cross-environment Wi-Fi HPE.
\end{abstract}


\section{Introduction}

\IEEEPARstart{3}{D} Human Pose Estimation (HPE) is an important enabling technology for smart environments, healthcare, and human--computer interaction. However, camera-based systems are sensitive to occlusion and lighting conditions and raise privacy concerns, while wearable solutions require intrusive on-body devices. Wi-Fi Channel State Information (CSI) has therefore emerged as a promising device-free and privacy-preserving sensing modality~\cite{yang2023mm}, with recent works demonstrating increasingly accurate and lightweight Wi-Fi HPE architectures~\cite{dao2026wiflowlightweightwifibasedcontinuous}. Despite this progress, reliable deployment remains challenging because Wi-Fi sensing models often generalize poorly to unseen environments.

The fundamental difficulty lies in the strong dependence of CSI on environmental multipath. Changes in room geometry and surrounding reflectors can substantially alter the observed CSI even for similar human poses, resulting in severe cross-environment domain shifts. Existing approaches address this problem through domain adaptation~\cite{10715589} or increasingly expressive feature-learning architectures~\cite{zhou2023metafi++}, yet robust HPE in unseen environments remains difficult. This motivates us to revisit what motion information is preserved in the CSI representation itself.

In particular, \emph{phase} provides a complementary but underexploited source of motion information. Existing Wi-Fi HPE methods often rely primarily on amplitude or collapse complex CSI into real-valued representations at an early stage~\cite{zhou2023metafi++}. Although raw CSI phase is corrupted by synchronization-induced offsets, properly sanitized phase variations retain motion-sensitive information associated with fine-grained propagation-path changes. Prematurely discarding these complex relationships may therefore remove useful motion cues before long-range temporal dynamics are learned. This leads to our central question: \emph{can complex phase relationships be preserved throughout spatio-temporal modeling to improve Wi-Fi HPE generalization across environments?}

To answer this question, we propose \textbf{C-MambaPose}, a phase-preserving complex-valued state-space framework for cross-environment Wi-Fi 3D HPE. Rather than using phase merely as an additional input, our key idea is to \emph{retain complex phase-amplitude relationships throughout sequence modeling}. C-MambaPose first mitigates synchronization-induced phase errors and employs Complex Dynamic Selective Kernel Convolution (Complex-DSKConv) to capture motion patterns at multiple scales. We then develop a Spatiotemporal Complex Mamba encoder that preserves complex-valued dynamics while efficiently modeling long-range dependencies using selective state-space models~\cite{10916584,10817504,djuhera2026mambacsphybridattentionstatespace}. Finally, cross-attention maps the learned sequence representations to joint-specific features, followed by a GraFormer decoder and bone-length regularization to exploit human skeletal structure.

Our main contributions are summarized as follows:
\begin{itemize}
    \item We develop a \textbf{phase-preserving complex CSI representation} that mitigates synchronization-induced phase errors while retaining motion-sensitive phase relationships, together with Complex-DSKConv for adaptive multi-scale feature extraction.

    \item We introduce a \textbf{Spatiotemporal Complex Mamba encoder} that preserves phase-amplitude relationships throughout selective state-space modeling, avoiding premature projection of complex CSI into real-valued features.

    \item We integrate \textbf{joint-query mapping with skeleton-aware graph decoding and kinematic regularization} for structurally coherent 3D HPE. Experiments on MM-Fi demonstrate strong cross-environment performance with only 3.78\,M parameters, while ablations show that preserving complex representations throughout sequence modeling is critical to the generalization gains.
\end{itemize}

\section{Methodology} \label{sec:method}
\begin{figure*}[tb]
    \centering
    \includegraphics[width=0.9\linewidth]{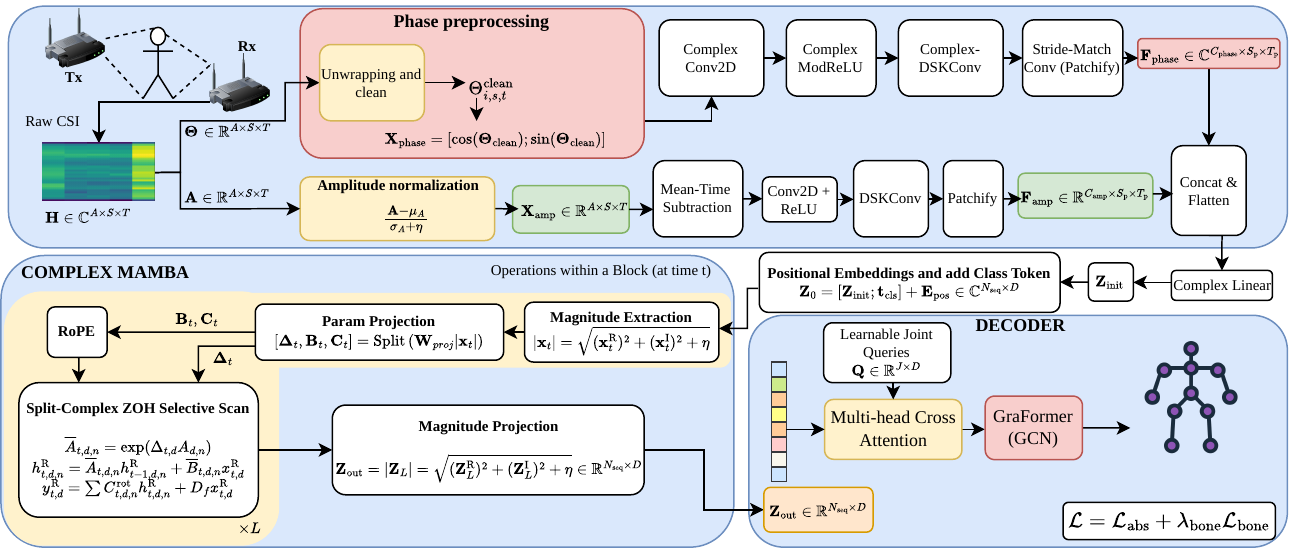}
    \vspace{-6pt}
    \caption{Overview of C-MambaPose framework.}
    \label{fig:framework}
\end{figure*}
C-MambaPose comprises three main stages, as illustrated in Fig.~\ref{fig:framework}: (1) CSI phase sanitization and decoupled formulation, (2) spatiotemporal Complex Mamba encoding with late fusion, and (3) cross-attention joint mapping with a regularized GraFormer decoder.

\subsection{CSI Phase Sanitization and Decoupled Formulation}

Let $\mathbf{H}\in\mathbb{C}^{A\times S\times T}$ denote the raw CSI, where $A$, $S$, and $T$ represent the numbers of receiving antennas, subcarriers, and temporal frames, respectively. Raw CSI phase contains synchronization-induced distortions arising from sampling and carrier frequency offsets. After unwrapping along the subcarrier dimension, the phase is modeled as
\begin{equation}
\theta^{\text{unwrapped}}_{i,s,t}
= s\,m_{i,t}+c_{i,t}
+\theta^{\text{clean}}_{i,s,t},
\label{eq:phase_model}
\end{equation}
where $m_{i,t}$ and $c_{i,t}$ denote the linear phase slope and offset, respectively. These terms are estimated by least-squares regression across subcarriers for each antenna and frame. The sanitized phase is then obtained as
\begin{equation}
\theta^{\text{clean}}_{i,s,t}
=\theta^{\text{unwrapped}}_{i,s,t}
-(s\,m_{i,t}+c_{i,t}).
\label{eq:phase_clean}
\end{equation}
To preserve phase coherence and raw angular relationships without early multipath interference, the phase is mapped onto the unit circle and represented as a concatenated real-valued tensor:
\begin{equation}
    \mathbf{X}_{\text{phase}} = \left[ \cos(\mathbf{\Theta}_{\text{clean}}) ; \sin(\mathbf{\Theta}_{\text{clean}}) \right] \in \mathbb{R}^{C_{\text{in}}^{\text{p}} \times S \times T}
\end{equation}
In parallel, the raw CSI amplitude $\mathbf{A} = |\mathbf{H}|$ is standardized sample-wise to eliminate static environmental domain mismatches:
\begin{equation}
    \mathbf{X}_{\text{amp}} = \frac{\mathbf{A} - \mu_{\mathbf{A}}}{\sigma_{\mathbf{A}} + \eta} \in \mathbb{R}^{C_{\text{in}}^{\text{a}} \times S \times T}
\end{equation}
where $\mu_{\mathbf{A}}$ and $\sigma_{\mathbf{A}}$ are the mean and standard deviation, and $\eta$ is a stabilizing constant.

\subsection{Spatiotemporal Complex Late-Fusion Mamba Encoder}

We propose a decoupled spatiotemporal encoder that extracts features from the phase and amplitude inputs using parallel branches, performs late fusion in the complex domain, and models long-range dynamics using a regularized Complex Mamba sequence processor.

\subsubsection{Parallel Feature Extraction}

The sanitized phase input $\mathbf{X}_{\text{phase}}$ and normalized amplitude input $\mathbf{X}_{\text{amp}}$ are processed by parallel feature extraction branches. 
For the \textbf{Phase Encoder Branch}, a complex-valued 2D convolution projects $\mathbf{X}_{\text{phase}}$ into a hidden space $\mathbf{Z}_{\text{phase}} \in \mathbb{C}^{C_{\text{phase}} \times S \times T}$. Non-linearity is introduced using the magnitude-dependent Complex-ModReLU activation~\cite{trabelsi2018deep}:
\begin{equation}
    \mathbf{U}_{\text{phase}} = \max(0, |\mathbf{Z}_{\text{phase}}| + \mathbf{b}_{\text{act}}^{\text{p}}) \odot \frac{\mathbf{Z}_{\text{phase}}}{|\mathbf{Z}_{\text{phase}}| + \eta}.
\end{equation}
We then apply a Complex Dynamic Selective Kernel Convolution (Complex-DSKConv) to adaptively adjust the receptive field. For $B$ parallel branches with different dilations yielding outputs $\mathbf{U}_b \in \mathbb{C}^{C_{\text{phase}} \times S \times T}$, we extract a joint magnitude envelope from their sum:
\begin{equation}
    |\mathbf{U}_{\text{sum}}| = \sqrt{\left(\sum_b \mathbf{U}_b^{\text{R}}\right)^2 + \left(\sum_b \mathbf{U}_b^{\text{I}}\right)^2 + \eta}.
\end{equation}
A global spatial channel-wise descriptor $\mathbf{s} \in \mathbb{R}^{C_{\text{phase}}}$ is subsequently computed via Global Average Pooling (GAP) on the magnitude envelope:
\begin{equation}
    s_c = \frac{1}{S \times T} \sum_{i=1}^{S} \sum_{t=1}^{T} |\mathbf{U}_{\text{sum}}|_{c, i, t}.
\end{equation}
A bottleneck projection is applied to generate a compact feature vector $\mathbf{z} = \text{ReLU}(\text{BN}(\mathbf{W}_1 \mathbf{s}))$. For each branch $b$, a soft attention weight vector $\alpha_b$ is derived using a softmax activation:
\begin{equation}
    \alpha_{b, c} = \frac{\exp(\mathbf{W}_{2,b,c} \mathbf{z})}{\sum_{k=1}^{B} \exp(\mathbf{W}_{2,k,c} \mathbf{z})}.
\end{equation}
The dynamically selected complex representation is computed as:
\begin{equation}
    \mathbf{U}_{\text{select}} = \sum_{b=1}^{B} \alpha_b \odot \mathbf{U}_b \in \mathbb{C}^{C_{\text{phase}} \times S \times T}
\end{equation}
which is followed by a convolution to generate phase patch tokens $\mathbf{F}_{\text{phase}}$. Parallelly, the \textbf{Amplitude Encoder Branch} dynamically normalizes $\mathbf{X}_{\text{amp}}$ over time, processes it via a real DSKConv, and patchifies it into $\mathbf{F}_{\text{amp}} \in \mathbb{R}^{C_{\text{amp}} \times S_{\text{p}} \times T_{\text{p}}}$.

\subsubsection{Late Fusion in Complex Space}

To merge the streams while preserving phase-amplitude relations, we project the real amplitude tokens to the complex domain ($\mathbf{F}'_{\text{amp}} = \mathbf{F}_{\text{amp}} + j \mathbf{0}$) and split the phase stream. The fused representation $\mathbf{F}_{\text{fused}}$ is constructed by concatenating components:
\begin{equation}
    \mathbf{F}_{\text{fused}} = \left[ \mathbf{F}_{\text{phase}}^{\text{R}} ; \mathbf{F}_{\text{amp}} \right] + j \left[ \mathbf{F}_{\text{phase}}^{\text{I}} ; \mathbf{0} \right].
\end{equation}
The tokens are flattened spatiotemporally into length $N_{\text{p}}$ and projected to embedding dimension $D$ via a complex-linear layer $\mathbf{Z}_{\text{init}}$. After Complex Layer Normalization, 1D complex positional embeddings $\mathbf{E}_{\text{pos}}$ and a class token $\mathbf{t}_{\text{cls}}$ are added:
\begin{equation}
    \mathbf{Z}_0 = \left[ \mathbf{Z}_{\text{init}} ; \mathbf{t}_{\text{cls}} \right] + \mathbf{E}_{\text{pos}} \in \mathbb{C}^{N_{\text{seq}} \times D}.
\end{equation}

\subsubsection{Sequence Processing with Regularized Complex Mamba}

The sequence $\mathbf{Z}_0$ is processed through $L$ layers of Spatiotemporal Complex Mamba blocks. For an input $\mathbf{x}_t \in \mathbb{C}^{D_{\text{inner}}}$, the selective parameters (step size $\mathbf{\Delta}_t$, and state projections $\mathbf{B}_t, \mathbf{C}_t$) are dynamically projected from the physical signal envelope:
\begin{equation}
    \left[ \mathbf{\Delta}_t, \mathbf{B}_t, \mathbf{C}_t \right] = \text{Split}\left(\mathbf{W}_{proj} \sqrt{(\mathbf{x}_t^{\text{R}})^2 + (\mathbf{x}_t^{\text{I}})^2 + \eta} \right).
\end{equation}

To explicitly inject relative spatiotemporal distance priors, Rotary Position Embeddings (RoPE) are applied directly to the time-varying state parameters $\mathbf{B}_t$ and $\mathbf{C}_t$. For a vector $\mathbf{v}_t$, adjacent elements are rotated at position $t$:
\begin{equation}
\begin{split}
    \mathbf{v}_{t, 2k}^{\text{rot}} &= \mathbf{v}_{t, 2k} \cos(t \theta_k) - \mathbf{v}_{t, 2k+1} \sin(t \theta_k), \\
    \mathbf{v}_{t, 2k+1}^{\text{rot}} &= \mathbf{v}_{t, 2k} \sin(t \theta_k) + \mathbf{v}_{t, 2k+1} \cos(t \theta_k)
\end{split}
\end{equation}
where $\theta_k = \theta_{\text{base}}^{-2k/D_{\text{state}}}$. The selective state matrix $\mathbf{A}$ and the position-encoded input matrix $\mathbf{B}_t^{\text{rot}}$ are discretized under the Zero-Order Hold (ZOH) assumption:
\begin{equation}
    \overline{A}_{t, d, n} = \exp(\Delta_{t, d} A_{d, n}),
\end{equation}
\begin{equation}
    \overline{B}_{t, d, n} = (\Delta_{t, d} A_{d, n})^{-1} (\exp(\Delta_{t, d} A_{d, n}) - 1) \Delta_{t, d} B_{t, d, n}^{\text{rot}}.
\end{equation}
Since the discretization parameters $\overline{A}_{t, d, n}$ and $\overline{B}_{t, d, n}$ are real-valued, the complex state transitions $h_{t, d, n} \in \mathbb{C}$ gracefully decouple into parallel real and imaginary paths:
\begin{equation}
\begin{split}
    h_{t, d, n}^{\text{R}} &= \overline{A}_{t, d, n} h_{t-1, d, n}^{\text{R}} + \overline{B}_{t, d, n} x_{t, d}^{\text{R}}, \\
    h_{t, d, n}^{\text{I}} &= \overline{A}_{t, d, n} h_{t-1, d, n}^{\text{I}} + \overline{B}_{t, d, n} x_{t, d}^{\text{I}}.
\end{split}
\end{equation}
The complex output $\mathbf{y}_t = \mathbf{y}_t^{\text{R}} + j \mathbf{y}_t^{\text{I}}$ is subsequently computed channel-wise as:
\begin{equation}
\begin{split}
    y_{t, d}^{\text{R}} &= \sum_{n=1}^{D_{\text{state}}} C_{t, d, n}^{\text{rot}} h_{t, d, n}^{\text{R}} + D_f x_{t, d}^{\text{R}}, \\
    y_{t, d}^{\text{I}} &= \sum_{n=1}^{D_{\text{state}}} C_{t, d, n}^{\text{rot}} h_{t, d, n}^{\text{I}} + D_f x_{t, d}^{\text{I}}.
\end{split}
\end{equation}
To execute this split-complex selective scan with high hardware efficiency, we stack inputs along the batch dimension $\mathbf{U}_{\text{stacked}} = [ \mathbf{x}^{\text{R}} ; \mathbf{x}^{\text{I}} ]$, allowing both trajectories to run in parallel in a single GPU kernel.

To prevent environmental overfitting, we apply a shared-mask Complex Dropout module:
\begin{equation}
    \text{ComplexDropout}(\mathbf{x}) = \frac{1}{1-p} \left[ \mathbf{x}^{\text{R}} \odot \mathbf{M}_{\text{drop}} ; \mathbf{x}^{\text{I}} \odot \mathbf{M}_{\text{drop}} \right].
\end{equation}
Using a shared mask $\mathbf{M}_{\text{drop}}$ ensures that phase relationships ($\angle \mathbf{x} = \arctan(\mathbf{x}^{\text{I}} / \mathbf{x}^{\text{R}})$) remain physically coherent. The sequence output is projected back to real-valued magnitude space: $\mathbf{Z}_{\text{out}} = |\mathbf{Z}_L| \in \mathbb{R}^{N_{\text{seq}} \times D}$.

\subsection{Cross-Attention Joint-Query Mapping}

To map unstructured tokens to the human skeleton, $J$ learnable joint-queries $\mathbf{Q} \in \mathbb{R}^{J \times D}$ scan the local spatiotemporal sequence tokens $\mathbf{M} = \mathbf{Z}_{\text{out}, 1:N_{\text{p}}}$ via Multi-Head Cross-Attention:
\begin{equation}
\begin{split}
    \mathbf{Z}_{\text{joint}} = \text{LN}\bigg(\mathbf{Q} &+ \text{softmax}\left(\frac{(\mathbf{Q}\mathbf{W}_Q)(\mathbf{M}\mathbf{W}_K)^T}{\sqrt{d_k}}\right) \\
    &\cdot \mathbf{M}\mathbf{W}_V\bigg).
\end{split}
\end{equation}

\subsection{Regularized GraFormer Decoder}

The features $\mathbf{Z}_{\text{joint}}$ are passed into a regularized Chebyshev Graph Convolutional Network (GraFormer~\cite{zhao2022graformer}) to predict coordinates while explicitly enforcing skeletal topology over graph $\mathcal{G} = (\mathcal{V}, \mathcal{E})$. A ChebGConv of order $K_{\text{cheb}}$ updates the features:
\begin{equation}
    \mathbf{Z}^{(l+1)} = \sum_{k=0}^{K_{\text{cheb}}-1} T_k(\tilde{\mathbf{L}}) \mathbf{Z}^{(l)} \mathbf{W}^{\text{GCN}}_k
\end{equation}
where $\tilde{\mathbf{L}}$ is the scaled normalized Laplacian, $T_k$ is the Chebyshev polynomial, and $\mathbf{W}^{\text{GCN}}_k$ are learnable weights. We propagate a strong dropout rate $p_{\text{gcn}}$ across all layers to prevent environmental overfitting and output coordinates $\hat{\mathbf{Y}} \in \mathbb{R}^{J \times D_{\text{coord}}}$.

\subsection{Training Loss Function}

The network is optimized using a multi-task loss function $\mathcal{L} = \mathcal{L}_{\text{abs}} + \lambda_{\text{bone}} \mathcal{L}_{\text{bone}}$. The \textbf{Absolute Position Loss ($\mathcal{L}_{\text{abs}}$)} penalizes Mean Squared Error (MSE) deviations of the predicted joints $\hat{\mathbf{Y}}_j$ from the ground-truth $\mathbf{Y}_j$:
\begin{equation}
    \mathcal{L}_{\text{abs}} = \frac{1}{J}\sum_{j=1}^{J} \|\hat{\mathbf{Y}}_j - \mathbf{Y}_j\|_2^2.
\end{equation}
Concurrently, the \textbf{Kinematic Bone Loss ($\mathcal{L}_{\text{bone}}$)} enforces anatomical constraints by minimizing the discrepancy between predicted and ground-truth bone lengths across the skeletal edges $\mathcal{E}$:
\begin{equation}
    \mathcal{L}_{\text{bone}} = \frac{1}{|\mathcal{E}|}\sum_{(u,v)\in\mathcal{E}} \left( \|\hat{\mathbf{Y}}_u - \hat{\mathbf{Y}}_v\|_2 - \|\mathbf{Y}_u - \mathbf{Y}_v\|_2 \right)^2.
\end{equation}
This combination explicitly prevents unnatural bone stretching and maintains skeletal topology under severe cross-environment domain shifts.

\section{Experiments}\label{sec:experiments}

\subsection{Experimental Setup}
\noindent\textbf{Dataset.} We evaluate the proposed C-MambaPose framework on the MM-Fi dataset~\cite{yang2023mm}. It contains 14 daily activities performed by 40 subjects across four distinct environments. The Wi-Fi system operates at 5 GHz with a 40 MHz bandwidth, consisting of one transmitter antenna and three receiver antennas. CSI is collected over 114 subcarriers, and consecutive frames are aggregated into samples of size $3 \times 114 \times 10$. Each skeletal pose is annotated with the 3D coordinates of 17 joints. Following the literature, we conduct our evaluations under Protocol 1 (focusing on the 14 daily activities) and Protocol 2 (focusing on the 13 rehabilitation exercises). To evaluate generalization, we follow three split settings:
\begin{itemize}
    \item \textbf{Setting 1 (Random Split - S1):} Data are randomly divided into training and testing sets with a 3:1 ratio.
    \item \textbf{Setting 2 (Cross-Subject Split - S2):} 32 subjects are used for training and the remaining 8 subjects for testing.
    \item \textbf{Setting 3 (Cross-Environment Split - S3):} Data from three environments are used for training and one for testing.
\end{itemize}

\noindent\textbf{Evaluation Metrics.} We adopt three standard metrics for 3D HPE: Mean Per Joint Position Error (MPJPE, in mm) to measure absolute joint distance; Procrustes Aligned MPJPE (PA-MPJPE, in mm) to measure structural pose accuracy; and Percentage of Correct Keypoints (PCK@20 and PCK@50) to report the ratio of joints predicted within a 20 mm or 50 mm threshold.

\noindent\textbf{Implementation Details.} Our model is implemented in PyTorch and trained on an NVIDIA RTX 4070 GPU. We use the AdamW optimizer with an initial learning rate of $3\times10^{-4}$ and a weight decay of 0.02. The learning rate is decayed to near zero over 50 epochs using a cosine annealing scheduler. The batch size is 256. The kinematic bone loss weight is configured as $\lambda_{\text{bone}}=0.1$ (while $\mathcal{L}_{\text{abs}}$ has a weight of 1.0). For reproducibility, we list the model hyperparameters explicitly: the Spatiotemporal Complex Mamba encoder employs $L=2$ complex SSM blocks with state dimension $d_{\text{state}}=16$, local convolution width $d_{\text{conv}}=4$, expansion factor $E=2$, and complex dropout rate $p=0.3$. The positional encoding is sinusoidal with a base frequency parameter $\theta_{\text{base}}=10,000.0$. The cross-attention joint-query mapper uses a learnable query dimension $d_{\text{emb}}=256$ with $4$ attention heads. The GCN decoder consists of a GraFormer with 4 graph layers, hidden dimension $d_{\text{hid}}=128$, and GCN dropout rate $p_{\text{gcn}}=0.4$. The Complex Dynamic Selective Kernel Convolution (Complex-DSKConv) operates with dilations $[1, 2]$. Our code is publicly available at \url{https://github.com/phucngvinuni/cmampose.git}.

\subsection{Comparison with State-of-the-Art Methods}
We compare C-MambaPose against existing state-of-the-art Wi-Fi-based 3D HPE baselines, including MetaFi++~\cite{zhou2023metafi++}, HPE-Li~\cite{d2024hpe}, and DT-Pose~\cite{chen2025towards}. Table~\ref{table:3d_mmfi} summarizes the comparisons across all three split settings.

\begin{table*}[t]
\caption{State-of-the-art comparisons on the MM-Fi dataset (Protocols 1 and 2). The best and second-best results are bolded and underlined, respectively.}
\vspace{-8pt}
\label{table:3d_mmfi}
\centering
\scalebox{0.82}{
\begin{tabular}{l|cccc|cccc}
\toprule
& \multicolumn{4}{c|}{Protocol 1 (Daily Actions)} & \multicolumn{4}{c}{Protocol 2 (Rehab Exercises)} \\
Method & PCK@20$\uparrow$ & PCK@50$\uparrow$ & MPJPE$\downarrow$ & PA-MPJPE$\downarrow$ & PCK@20$\uparrow$ & PCK@50$\uparrow$ & MPJPE$\downarrow$ & PA-MPJPE$\downarrow$ \\
\midrule
\multicolumn{9}{l}{\textit{Setting 1 (Random Split):}}\\
MetaFi++~\cite{zhou2023metafi++}      & 49.1 & 86.5 & 186.9 & 120.7 & 32.2 & 81.7 & 213.5 & 121.4 \\
HPE-Li~\cite{d2024hpe}                & 56.2 & 87.6 & 173.4 & 104.5 & 36.9 & 81.9 & 206.1 & 102.7 \\
GraphPose-Fi~\cite{chen2026graph}     & \textbf{61.1} & \textbf{89.3} & \textbf{160.6} & 105.0 & 41.0 & \textbf{84.3} & \textbf{193.5} & 103.3 \\
DT-Pose~\cite{chen2025towards}        & 59.4 & 88.9 & 165.3 & \underline{101.0} & \underline{41.4} & 83.5 & 195.6 & \underline{101.2} \\
\textbf{Ours (C-MambaPose)}            & \underline{59.6} & \underline{89.1} & \underline{165.1} & \textbf{100.8} & \textbf{41.6} & \underline{83.8} & \underline{194.8} & \textbf{99.8} \\
\midrule
\multicolumn{9}{l}{\textit{Setting 2 (Cross-Subject Split):}}\\
MetaFi++~\cite{zhou2023metafi++}      & 36.4 & 85.5 & 222.3 & 125.4 & 24.0 & 77.5 & 247.0 & 122.7 \\
HPE-Li~\cite{d2024hpe}                & 38.2 & 82.8 & 228.6 & 106.8 & 26.9 & 78.0 & 242.6 & 101.9 \\
GraphPose-Fi~\cite{chen2026graph}     & \textbf{44.2} & 86.3 & \textbf{210.5} & \underline{105.5} & \textbf{31.8} & \textbf{80.5} & \textbf{227.6} & 102.8 \\
DT-Pose~\cite{chen2025towards}        & 41.9 & \underline{86.7} & 213.0 & 105.6 & 28.5 & 78.5 & 238.3 & \underline{101.1} \\
\textbf{Ours (C-MambaPose)}            & \underline{42.2} & \textbf{87.1} & \underline{211.5} & \textbf{102.8} & \underline{28.8} & \underline{78.8} & \underline{236.8} & \textbf{100.8} \\
\midrule
\multicolumn{9}{l}{\textit{Setting 3 (Cross-Environment Split):}}\\
MetaFi++~\cite{zhou2023metafi++}      & 9.3 & 55.1 & 367.8 & 121.0 & \underline{5.3} & 45.9 & 360.2 & 117.2 \\
HPE-Li~\cite{d2024hpe}                & 4.3 & 47.8 & 381.1 & 110.3 & 4.2 & 40.3 & 378.2 & 104.0 \\
GraphPose-Fi~\cite{chen2026graph}     & \underline{12.9} & \textbf{67.2} & \underline{302.7} & \underline{103.0} & \underline{5.3} & 42.0 & 373.4 & 105.4 \\
DT-Pose~\cite{chen2025towards}        & 10.7 & 58.8 & 332.7 & 105.1 & 4.4 & \underline{49.7} & \underline{338.3} & \underline{102.0} \\
\textbf{Ours (C-MambaPose)}            & \textbf{15.3} & \underline{67.1} & \textbf{298.5} & \textbf{102.2} & \textbf{8.4} & \textbf{56.9} & \textbf{317.6} & \textbf{101.1} \\
\bottomrule
\end{tabular}
}
\vspace{-10pt}
\end{table*}

\noindent\textbf{Quantitative Results.} As shown in Table~\ref{table:3d_mmfi}, C-MambaPose exhibits highly competitive performance under Setting 1 (Random Split) and Setting 2 (Cross-Subject), performing on par with existing state-of-the-art architectures. Specifically, under Setting 1, our model achieves a PA-MPJPE of \textbf{100.8 mm} (matching or slightly outperforming DT-Pose's 101.0 mm). Under Setting 2, our model achieves a PA-MPJPE of \textbf{102.8 mm} (which is comparable to the best baseline performance).

\begin{figure*}[tb]
    \centering
    \includegraphics[width=0.95\linewidth]{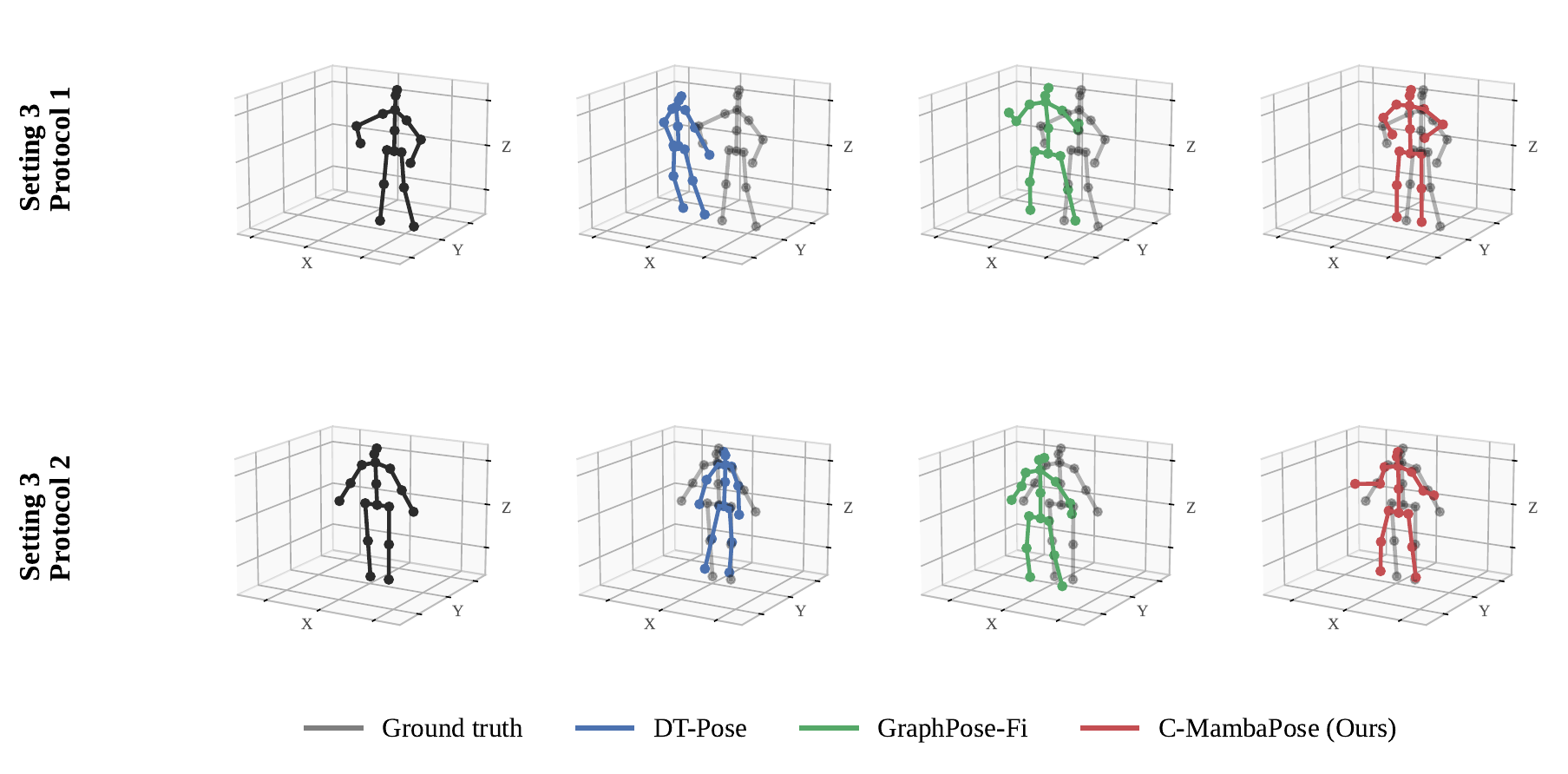}
    \caption{Qualitative visualization of 3D human pose estimation on the MM-Fi dataset under the challenging Setting 3 (Cross-Environment split). The top row illustrates an activity from Protocol 1, while the bottom row shows a rehabilitation exercise from Protocol 2.}
    \label{fig:qualitative}
\end{figure*}

Crucially, in the most challenging cross-environment split (Setting 3) where severe domain shifts occur across rooms, our model achieves significant improvements. C-MambaPose reduces the MPJPE to \textbf{298.5 mm} (outperforming GraphPose-Fi by 4.2 mm and DT-Pose by 34.2 mm) and achieves a competitive PCK@50 of \textbf{67.1\%} and a state-of-the-art PCK@20 of \textbf{15.3\%}. These improvements confirm that preserving complex phase dynamics via our Spatiotemporal Mamba and GraFormer structure successfully prevents overfitting to specific environmental layouts. As visually demonstrated in Fig.~\ref{fig:qualitative}, baselines suffer from severe positional drift and skeleton distortion in unseen rooms,

We also observe an interesting stability pattern across the split settings: while the absolute joint coordinate error (MPJPE) degrades significantly under severe domain shift in Setting 3 (rising from 165.1 mm in S1 to 298.5 mm in S3), the Procrustes Aligned error (PA-MPJPE) remains remarkably stable (100.8 mm in S1 to 102.2 mm in S3). This indicates that while environmental multipath noise causes shifts in absolute localization coordinates, the relative skeletal structure and relative joint relationships are preserved. This supports our thesis that integrating a GCN GraFormer decoder enforces structural skeleton connectivity priors, ensuring anatomically valid pose predictions even under severe environmental shift.

\noindent\textbf{Model Complexity and Runtime Benchmarks.} 
Table~\ref{table:complexity} compares the parameter sizes, inference latency, and localization performance of different models on Setting 3.

\begin{table}[H]
\caption{Model complexity, inference latency, and performance comparisons on MM-Fi (Setting 3). The best and second-best results in each category are bolded and underlined, respectively.}
\vspace{-8pt}
\label{table:complexity}
\centering
\scalebox{0.85}{
\begin{tabular}{l|c|c|cc}
\toprule
Model Architecture & Params (M) $\downarrow$ & Latency (ms) $\downarrow$ & MPJPE $\downarrow$ & PCK@50 $\uparrow$ \\
\midrule
MetaFi++~\cite{zhou2023metafi++}      & 26.52 & 59.94 & 367.8 & 55.1 \\
GraphPose-Fi~\cite{chen2026graph}     & 22.43 & 62.61 & \underline{302.7} & \textbf{67.2} \\
DT-Pose~\cite{chen2025towards}        & \underline{3.20} & \underline{4.23} & 332.7 & 58.8 \\
HPE-Li~\cite{d2024hpe}                & \textbf{0.86} & \textbf{2.24} & 381.1 & 47.8 \\
\textbf{Ours}            & 3.78 & 25.43 & \textbf{298.5} & \underline{67.1} \\
\bottomrule
\end{tabular}
}
\vspace{-10pt}
\end{table}

Our proposed C-MambaPose achieves a highly competitive trade-off between model capacity, runtime efficiency, and localization performance. Specifically, C-MambaPose requires a highly compact parameter footprint (3.78 M parameters)-an 83.1\% reduction compared to GraphPose-Fi~\cite{chen2026graph} (22.43 M) and is comparable in size to DT-Pose~\cite{chen2025towards} (3.20 M, which is only 18\% smaller), while achieving significantly cross-environment performance (298.5 mm MPJPE on Setting 3, compared to 302.7 mm for GraphPose-Fi and 332.7 mm for DT-Pose) without requiring any pretraining. 

\subsection{Ablation Study}
To evaluate the effectiveness of the individual components of C-MambaPose, we conduct comprehensive ablation studies on Setting 3 (Cross-Environment) of the MM-Fi dataset.

\noindent\textbf{Architecture Components.}
Table~\ref{table:ablation_arch} analyzes the impact of the complex-valued formulation, Complex-DSKConv, GraFormer decoder, the placement of the magnitude projection bottleneck, the Rotary Position Embedding (RoPE), and the Shared-Mask Complex Dropout.
\begin{table}[H]
\caption{Ablation study of model architectural components on MM-Fi (Setting 3).}
\vspace{-8pt}
\label{table:ablation_arch}
\centering
\scalebox{0.85}{
\begin{tabular}{l|cc}
\toprule
Configuration & MPJPE (mm) $\downarrow$ & PA-MPJPE (mm) $\downarrow$ \\
\midrule
Ours (Full Model)             & \textbf{298.5} & \textbf{102.2} \\
w/o Complex Representation     & 365.5 & 103.5 \\
w/o Complex-DSKConv           & 339.5 & 103.9 \\
w/o GraFormer (MLP decoder)   & 358.9 & 103.2 \\
Phase Bottleneck (Mag before Mamba) & 335.0 & 102.3 \\
w/o Rotary Position Embedding (RoPE) & 312.4 & 102.9 \\
w/o Shared-Mask Complex Dropout      & 321.8 & 103.1 \\
\bottomrule
\end{tabular}
}
\vspace{-5pt}
\end{table}

Replacing our complex-valued pipeline with a standard real-valued amplitude-only input leads to a degraded MPJPE of 365.5 mm, highlighting that phase information is crucial for tracking Doppler shifts in unseen domains. Furthermore, removing the Complex-DSKConv module yields a degraded MPJPE of 339.5 mm, proving that a dynamic multi-scale receptive field is essential to capture varied motion speeds. Replacing the GraFormer GCN with an MLP decoder degrades performance to 358.9 mm, demonstrating the regularizing effect of joint connectivity priors. Projecting complex features to real magnitude before sequential processing (Phase Bottleneck) leads to a degraded MPJPE of 335.0 mm, validating that preserving phase relations inside the Mamba sequence blocks is vital.

Additionally, removing the Rotary Position Embedding (RoPE) from state transitions degrades the MPJPE to 312.4 mm, confirming that explicitly encoding relative spatiotemporal token distances is crucial for tracking sequential body movements. Finally, removing the Shared-Mask Complex Dropout (i.e., applying independent dropout masks to real and imaginary channels) degrades the MPJPE to 321.8 mm. This is because independent masks disrupt the phase angle relationships ($\angle \mathbf{x} = \arctan(\mathbf{x}^{\text{I}} / \mathbf{x}^{\text{R}})$) between the real and imaginary channels, introducing random phase distortion that degrades physical signal coherence and generalization.

\noindent\textbf{Impact of Kinematic Bone Loss.}
Table~\ref{table:ablation_loss} evaluates the model under different values of the kinematic bone loss coefficient $\lambda_{\text{bone}}$ when added to the absolute position loss baseline $\mathcal{L} = \mathcal{L}_{\text{abs}} + \lambda_{\text{bone}} \mathcal{L}_{\text{bone}}$.
\begin{table}[H]
\caption{Ablation of kinematic loss coefficient $\lambda_{\text{bone}}$ on MM-Fi (Setting 3).}
\vspace{-8pt}
\label{table:ablation_loss}
\centering
\scalebox{0.88}{
\begin{tabular}{l|cc}
\toprule
Loss Configuration & MPJPE (mm) $\downarrow$ & PA-MPJPE (mm) $\downarrow$ \\
\midrule
$\lambda_{\text{bone}} = 0$ (MPJPE Loss only) & 314.3 & 102.4 \\
$\lambda_{\text{bone}} = 0.1$ (Ours)           & \textbf{298.5} & \textbf{102.2} \\
$\lambda_{\text{bone}} = 0.5$                  & 302.5 & 103.0 \\
\bottomrule
\end{tabular}
}
\vspace{-10pt}
\end{table}

Without the bone length constraint ($\lambda_{\text{bone}} = 0$), the model achieves 314.3 mm MPJPE. Introducing the constraint with $\lambda_{\text{bone}} = 0.1$ guides the model to maintain physical skeletal symmetry, yielding the best results. However, setting the weight too high ($\lambda_{\text{bone}} = 0.5$) overly penalizes joint movements and restricts positional flexibility, worsening the errors to 302.5 mm.

\section{Conclusion and Discussion}\label{sec:conclusion}
In this paper, we presented C-MambaPose, a phase-preserving complex-valued state-space framework for robust cross-environment Wi-Fi 3D HPE. By preserving complex phase-amplitude relationships throughout spatio-temporal modeling and incorporating skeletal priors, C-MambaPose effectively captures motion dynamics while maintaining anatomically coherent poses. Experiments on MM-Fi demonstrate strong cross-environment performance with only 3.78\,M parameters, while ablations highlight the importance of retaining complex representations throughout sequence modeling. Future work will extend C-MambaPose to multi-person sensing and explore domain adaptation for further improving generalization to unseen environments.

\section*{Acknowledgement}
This work was supported by the VinUniversity project CECS.UNI.2026.DA012

\bibliographystyle{IEEEtran}
\bibliography{refs.bib}

@ARTICLE{zhou2023metafi++,
  author={Zhou, Yunjiao and Huang, He and Yuan, Shenghai and Zou, Han and Xie, Lihua and Yang, Jianfei},
  journal={IEEE Internet of Things Journal}, 
  title={MetaFi++: WiFi-Enabled Transformer-Based Human Pose Estimation for Metaverse Avatar Simulation},
  doi={10.1109/JIOT.2023.3262940}}

@InProceedings{d2024hpe,
author="D. Gian, Toan
and Dac Lai, Tien
and Van Luong, Thien
and Wong, Kok-Seng
and Nguyen, Van-Dinh",
editor="Leonardis, Ale{\v{s}}
and Ricci, Elisa
and Roth, Stefan
and Russakovsky, Olga
and Sattler, Torsten
and Varol, G{\"u}l",
title="HPE-Li: WiFi-Enabled Lightweight Dual Selective Kernel Convolution for Human Pose Estimation",
booktitle="Computer Vision -- ECCV 2024",
year="2025",
publisher="Springer Nature Switzerland",
address="Cham",
pages="93--111",
isbn="978-3-031-72751-1"
}

@inproceedings{
      yang2023mm,
      title={MM-Fi: Multi-Modal Non-Intrusive 4D Human Dataset for Versatile Wireless Sensing},
      author={Yang, Jianfei and Huang, He and Zhou, Yunjiao and Chen, Xinyan and Xu, Yuecong 
              and Yuan, Shenghai and Zou, Han and Lu, Chris Xiaoxuan and Xie, Lihua},
      booktitle={Thirty-seventh Conference on Neural Information Processing Systems Datasets and Benchmarks Track},
      year={2023},
  }

@inproceedings{zhao2022graformer,
  title={Graformer: Graph-oriented transformer for 3d pose estimation},
  author={Zhao, Weixi and Wang, Weiqiang and Tian, Yunjie},
  booktitle={Proceedings of the IEEE/CVF conference on computer vision and pattern recognition},
  pages={20438--20447},
  year={2022}
}

@inproceedings{chen2026graph,
  title={Graph-Based 3d Human Pose Estimation Using Wifi Signals},
  author={Chen, Jichao and Qu, YangYang and Tang, Ruibo and Slock, Dirk},
  booktitle={ICASSP 2026-2026 IEEE International Conference on Acoustics, Speech and Signal Processing (ICASSP)},
  pages={19992--19996},
  year={2026},
  organization={IEEE},
doi = {10.1109/ICASSP55912.2026.11463884}
}

@ARTICLE{10715589,
  author={Zheng, Naiyu and Li, Yuanchun and Jiang, Shiqi and Li, Yuanzhe and Yao, Rongchun and Dong, Chuchu and Chen, Ting and Yang, Yubo and Yin, Zhimeng and Liu, Yunxin},
  journal={IEEE Transactions on Mobile Computing}, 
  title={AdaWiFi, Collaborative WiFi Sensing for Cross-Environment Adaptation}, 
  year={2025},
  volume={24},
  number={2},
  pages={845-858},
  doi={10.1109/TMC.2024.3474853}}

@ARTICLE{10916584,
  author={Huang, Yunlong and Liu, Junshuo and Shi, Xin and Zhao, Sijie and Mi, Tiebin and Qiu, Robert C.},
  journal={IEEE Sensors Journal}, 
  title={SenseMamba: A General Lightweight State-Space Model for Wireless Human Sensing}, 
  year={2025},
  volume={25},
  number={8},
  pages={13859-13870},
  doi={10.1109/JSEN.2025.3546619}}

@ARTICLE{10817504,
  author={Liu, Junshuo and Huang, Yunlong and Shi, Xin and Ren, Xiang and Mi, Tiebin and Qiu, Robert C.},
  journal={IEEE Sensors Journal}, 
  title={TF-Mamba: A Lightweight State-Space Model for Wi-Fi-Based Human Activity Recognition}, 
  year={2025},
  volume={25},
  number={13},
  pages={23184-23194},
  doi={10.1109/JSEN.2024.3520857}}

@misc{djuhera2026mambacsphybridattentionstatespace,
      title={MambaCSP: Hybrid-Attention State Space Models for Hardware-Efficient Channel State Prediction}, 
      author={Aladin Djuhera and Haris Gacanin and Holger Boche},
      year={2026} 
}

@misc{dao2026wiflowlightweightwifibasedcontinuous,
      title={WiFlow: A Lightweight WiFi-based Continuous Human Pose Estimation Network with Spatio-Temporal Feature Decoupling}, 
      author={Yi Dao and Lankai Zhang and Hao Liu and Haiwei Zhang and Wenbo Wang},
      year={2026},
      eprint={2602.08661},
      archivePrefix={arXiv} 
}

@inproceedings{
trabelsi2018deep,
title={Deep Complex Networks},
author={Chiheb Trabelsi and Olexa Bilaniuk and Ying Zhang and Dmitriy Serdyuk and Sandeep Subramanian and Joao Felipe Santos and Soroush Mehri and Negar Rostamzadeh and Yoshua Bengio and Christopher J Pal},
booktitle={International Conference on Learning Representations},
year={2018}
}

@misc{chen2025towards,
      title={Towards Robust and Realistic Human Pose Estimation via WiFi Signals}, 
      author={Yang Chen and Jingcai Guo},
      year={2026},
      eprint={2501.09411},
      archivePrefix={arXiv},
      primaryClass={cs.CV},
      url={https://arxiv.org/abs/2501.09411}, 
}

\end{document}